# The role of geographic proximity in knowledge diffusion, measured by citations to scientific literature[1]

Giovanni Abramo

*Laboratory for Studies in Research Evaluation*
*at the Institute for System Analysis and Computer Science (IASI-CNR)*
*National Research Council of Italy*
ADDRESS: Istituto di Analisi dei Sistemi e Informatica, Consiglio Nazionale delle Ricerche, Via dei Taurini 19, 00185 Roma - ITALY
giovanni.abramo@uniroma2.it

Ciriaco Andrea D'Angelo

*University of Rome "Tor Vergata" - Italy and*
*Laboratory for Studies in Research Evaluation (IASI-CNR)*
ADDRESS: Dipartimento di Ingegneria dell'Impresa, Università degli Studi di Roma "Tor Vergata", Via del Politecnico 1, 00133 Roma - ITALY
dangelo@dii.uniroma2.it

Flavia Di Costa

*Research Value s.r.l.*
ADDRESS: Research Value, Via Michelangelo Tilli 39, 00156 Roma- ITALY
flavia.dicosta@gmail.com

**Abstract**
This paper analyses the influence of geographic distance on knowledge flows, measured through citations to scientific publications. Previous works using the same approach are limited to single disciplines. In this study, we analyse the Italian scientific production in all disciplines matured in the period 2010-2012. To calculate the geographic distances between citing and cited publications, each one is associated with a "prevalent" territory on the basis of the authors' affiliations. The results of the application of a gravity model, estimated using ordinary least squares regression, show that despite the spread of IT, geographic distance continues to be an influential factor in the process of knowledge flows between territories. In particular, the analysis reveals that the effect of geographic distance on knowledge flows is significant at the national level, not negligible at the continental level, but completely irrelevant at the intercontinental level.

# 1. Introduction

Scholars and policy makers agree that knowledge is one of the key drivers of long-run economic growth. Scientific and technological progress is achieved through a continuous process of knowledge transfer within and between communities of scientists and technologists. The speed and breadth of knowledge diffusion influence the intensity of progress. The development of information technologies in recent decades has contributed significantly to breaking down geographic barriers, promoting the spread of knowledge and reducing the time (Ding, Levin, Stephan, & Winkler, 2010).

The questions of how knowledge spreads and how this can be tracked have attracted the attention of many scholars. The recent development of bibliometric techniques has favoured investigation into knowledge flows, both in terms of the breadth of the field that can be observed and for in-depth analysis of aspects of the phenomenon. Recently, Abramo and D'Angelo (2018) traced the international spillovers of knowledge produced in Italy, in more than 200 fields. Subsequently, Abramo, D'Angelo, and Carloni (2019) first conceptualized the "balance of knowledge flows" (BKF), registering the flows of knowledge across countries, and paralleling the concept of technological balance of payments (TBP). Among other aspects, the authors measured the share of domestic versus foreign flows generated by a country's research system, by field and as compared to other countries. Finally, operating at the NUTS2[2] geographic level, the same authors applied the concept of BKF to measure the spillover of knowledge across the Italian regions (Abramo & D'Angelo, 2019).

From the outset, scholars have been curious about the relation of knowledge flows to the geographic distance between producer and user. Indeed, it has been stated that "while knowledge has a high propensity to spillover, such knowledge spillovers are geographically bounded" (Audretsch & Lehmann, 2005; Audretsch & Keilbach, 2007). In this regard, although the costs and time for transacting information between economic agents have been progressively reduced, especially for encoded information, knowledge also has a significant component of tacit understanding. The transfer of such tacit knowledge is expensive and requires face-to-face communication, meaning that the "geographic factor" remains important and not negligible.

Almost all studies on the effects of geographic proximity on knowledge flows are based on the analysis of patent citations. Relatively few have analyzed article citations, and these in turn have focused on specific sectors.

The current study enters the research stream on effects of geographic proximity on knowledge flows, however it overcomes the sectoral limit by applying a survey that covers the entire scientific spectrum. Like the previous ones, it starts from the assumption that citation linkages between articles imply a flow of knowledge from the cited to the citing authors (Mehta, Rysman, & Simcoe, 2010; Van Leeuwen & Tijssen, 2000). It aims, therefore, to study the influence of the variable of "geographic distance" on the citation flows between "cited" and "citing" publications.

The dataset for the analysis consists of the Italian scientific publications indexed in the Web of Science (WoS) core collection in the period 2010-2012 and the related citations received as of the close of 2017. Each publication (citing and cited) was associated with a prevalent territory in order to address the following research questions:

---

[2] Nomenclature of Territorial Units for Statistics or NUTS is a geocode standard for referencing the subdivisions of countries for statistical purposes.

- What is the weight of the "distance" factor on the knowledge spillover generated from the scientific publications produced in the different territories?
- How does the weight of this factor vary with respect to the geographic dimension considered (i.e. national, continental, intercontinental)?

The study differs from the previous ones in size and breadth of field of study. The dataset and methods applied make it is possible to observe the flows generated by all WoS-indexed Italian publications over a three-year period, without any kind of restriction by scientific field - meaning 161,680 publications, cited by 1,800,037.

The next section summarizes the literature on the subject with particular emphasis on methodology. Section 3 presents the data and analytical method, and Section 4 the results from the elaborations. Section 5 closes the work with a synthesis of main results and the authors' considerations of the implications.

## 2. Literature review

Many scholars have investigated the phenomenon of knowledge diffusion, also referred as "knowledge spillovers" and "knowledge flows" (Jaffe, Trajtenberg, & Henderson, 1993; Hicks, Breitzman, Olivastro & Hamilton, 2001; Tijssen, 2001; Maurseth & Verspagen, 2002; Thompson & Fox-Kean, 2005; Sonn & Storper, 2008; Belenzon & Schankerman, 2013; Jaffe & De Rassenfosse, 2017; Abramo, D'Angelo, & Carloni, 2019). A remarkable work of classification of the large amount of studies on the subject was carried out by Cerver-Romero, Ferreira and Fernandes (2018).

From a methodological point of view, the literature seems to propose two distinct bibliometric approaches: one uses social network analysis applied to co-authorship networks (Capello & Caragliu, Newman, 2003, Yang & Ding, 2012; Yan, Ding & Kong, 2012); the other uses citing-cited relations in the scientific literature.

From our point of view, the first approach is more suitable in studying the exchange of tacit knowledge for creation of new knowledge (including integration of different competences), while the second is more appropriate to study the exploitation of knowledge actually produced (no longer tacit, made public) to create new knowledge. This second aspect is the one that interests us more, and which we explore in this study.

In the latter area, the studies differ in the type of knowledge investigated, i.e. that encoded in the patent literature and that in the scientific literature.

In the application of citation analysis to the study of knowledge spillovers encoded in patent literature, one of the milestones is the work of Jaffe, Trajtenberg and Henderson (1993), whose underlying rationale is that "knowledge flows do sometimes leave a paper trail, in the form of citations in patents". Over the years further studies have explored various aspects of the phenomenon (Hicks, Breitzman, Olivastro & Hamilton, 2001; Tijssen, 2001; Maurseth & Verspagen, 2002; Thompson & Fox-Kean, 2005; Sonn & Storper, 2008; Belenzon & Schankerman, 2013; Jaffe & De Rassenfosse, 2017).

In the seminal article of Jaffe, Henderson, and Trajtenberg (1993), the comparison concerns the geographic location of patent citations with that of the cited patents, to assess whether knowledge spillovers are geographically localized. The field of observation consists of two cohorts of American patents, 1400 in all. The analysis was carried out at the level of country, state, and municipality assigned on the basis of the plurality of the authors. Findings show that citations to domestic patents are more likely to be domestic, and more likely to come from the same state and municipality.

More recently, Belenzon and Schankerman (2013) confirmed that the diffusion of knowledge produced in American universities is influenced by factors such as state borders and distance. To identify knowledge flow trajectories, they used patent citations both to university patents and scientific publications. The results show that the citations decline drastically with increasing range, up to 150 miles (241 km), and then remain constant for longer distances. The presence of a threshold value corresponding to an extended commuting distance (working distance) indicates that the existence of direct personal interaction plays an important role in knowledge flows.

A number of works have investigated the flows of knowledge encoded in the scientific literature. They are all focused on specific scientific fields though, and results are not always unequivocal. A methodological contribution to the topic is offered by Frenken, Hardeman, and Hoekman (2009) who proposed an analytical framework able to distinguish between physical proximity and other forms of "proximity" as determinants of scientific interaction.

Matthiessen, Schwarz, and Find (2002) analyzed networks of research co-operation between 40 largest urban regions using publications indexed in the Science Citation Index 1997-1999. They found that the citing patterns are very weakly dependent on distance, but not of nationality. Conversely, using a 20-year PNAS publication data of major U.S. Research Institutions, Börner, Penurnarthy, Meiss, and Ke (2006) observed a strong log-linear relationship between citations exchanged by Institutions and their geographical distance, concluding that "the citation linkages between institutions fall off with the distance between them". A similar conclusion is reached by Pan, Kaski, and Fortunato (2012). Analyzing all publications indexed in the WoS in 2003–2010 period they claimed that the citation flows as well as the collaboration strengths between cities decrease with the distance between them, following a gravity law.

Ahlgren, Persson and Tijssen (2013) analyzed citation-based relations in publications appearing in the journal Scientometrics from 1981 to 2010, using different measures of mean geographic distance (MGD). This study also concludes that the local effect remains an important factor in knowledge flows. Yan & Sugimoto (2011) analyzed citation flows and collaboration networks between institutions in the library and information science field, and claimed that scholars tend to cite colleagues of the same country and/or those physically close to them. However, the steady introduction of online databases has weakened the effect of the physical distance so that the citations have become more closely dependent on the intensity of collaboration.

Recently, Wuestman, Hoekman, and Frenken (2019) investigated the geography of 2014 articles in the life sciences and medicine citing those published in 2012. They questioned what they call "geographical bias in citations", claiming that self-citations are an important driver of such bias. Moreover, once "cognitive relatedness" (measured by the number of references shared by two publications) is accounted for, the effect of distance between citing and cited publications is weak. The authors warn about the generalizability of their findings due to the sector and time specific analysis they have conducted. Also Head, Li, and Minondo (2018) conclude that the negative impact of geographic distance on citations is "mediated" by "social relatedness". They studied how geographic distance and social ties (co-authorship, past collocation, and relationships mediated by advisors and the alma mater) affect citation patterns in mathematics, observing that when controlling for ties, the negative impact of geographic distance on citations is generally halved. The authors hypothesize that spatial proximity facilitates the creation of interpersonal links that in turn favor knowledge flows.

## 3. Method and data

Different explanatory models have been applied in the attempt to study the geographic factor in knowledge flows, the most famous being "gravity models". The gravitational model theorized by Tinbergen (1962) originated in the economic field and then found widespread application in the study of international economics phenomena, such as the extent of bilateral trade between countries (Anderson, 1979; Anderson, 2011; Deardoff, 1998; Anderson & Van Wincoop, 2003; Brakman & van Marrewijk, 2008). More recently, these models have been developed for new contexts, such as for the study of scientific collaborations between different types of institutions (Ponds, Van Oort, & Frenken, 2007).

To test the influence of geographic distance on knowledge flows we will again use a gravitational model, based on two assumptions:
- the flow of knowledge between any two territories can be measured through the citations made in the scientific production by the research centres in the first territory, to the scientific production by the research centres in the second (i.e. citations in the scientific literature of the "citing territory" to the scientific literature of the "cited territory");
- citations between two territories increase with the amount of scientific production of both, and decrease with the distance between them.

From an operational point of view, the preparatory work for the elaboration required three steps: i) construction of the dataset, consisting of the pairs of cited and citing publications; ii) assignment of the geographic attribute to each cited publication and the relevant citing ones; iii) calculation of the geographic distances between citing and cited publications.

The Clarivate Analytics Italian national citation report (I-NCR) for 2010-2012 registers all publications with "Italy" in the affiliation list. Let P denote the set of the cited publications indexed in such report. For each publication in P, we reduce all addresses to city + country expressions (e.g. "Rome, Italy"). Each "city" is then matched to the corresponding LAU level (local administrative unit, 7915 in all)[3] called *comune* in Italy, using the official lists of the National Institute of Statistics (ISTAT).[4]

Subsequently, it was necessary to attribute the publications to the "prevalent" territory, enabling measurement of the flows of knowledge between the territories of production, given that these would then be identified for both cited and citing publications. However, because of increasing research collaboration at both national and international levels (Uddin, Hossain, Abbasi, & Rasmussen, 2012; Larivière, Gingras, Sugimoto, & Tsou, 2015), identifying the territory of production of a publication can be quite complex. Various approaches could be envisaged to assign the publications to geographic entities:
  i)   to each of the territories of the institutions in the address list;
  ii)  to one single territory, by the frequency of authors (or institutions) of the territory in the address list, or by the affiliation of the corresponding author, or by the affiliation of the first and last authors in non-alphabetically ordered bylines;
  iii) by fractionalizing the publication by the number of territories, institutions or authors.

We determined to adopt two distinct conventions for the cited and citing publications:

---

[3] The LAU level consists of municipalities or equivalent units in the 28 EU Member States.
[4] https://www.istat.it/it/archivio/6789, last access 8 January, 2020.

Cited publications: Since I-NCR contains the affiliation of each author, we define a publication as "made in" an Italian territory if the greatest share of co-authors are affiliated to organizations located in that territory. To exemplify, consider the publication with DOI 10.3389/fpsyg.2011.00227, whose byline is shown in the box below:

---

Scorolli, C[1]; Binkofski, F[2]; Buccino, G[3]; Nicoletti, R[4]; Riggio, L[5]; Borghi, AM[1,6]
   [1] Univ Bologna, Dept Psychol, I-40127 Bologna, Italy
   [2] Rhein Westfal TH Aachen, Div Cognit Neurol, D-52062 Aachen, Germany
   [3] Univ Catanzaro, Dept Med Sci, Catanzaro, Italy
   [4] Univ Bologna, Dept Commun Disciplines, Bologna, Italy
   [5] Univ Parma, Dept Neurosci, I-43100 Parma, Italy
   [6] CNR, Inst Cognit Sci & Technol, Rome, Italy

---

Fractionalizing the authorship of "Borghi, AM" (who shows two distinct affiliations) and applying the above convention, we assign the publication to the city of "Bologna", totaling 2.5/6=41.7% of authorships.

Citing publications: differently from the cited publications, for the citing publications the I-NCR reports only the address list without the link to authors. We define then a citing publication as "made in" a territory if the greatest share of addresses refer to that territory.[5] To exemplify, consider the publication with DOI 10.1182/blood-2010-01-261289, whose address list is:

---

   Catholic Univ Korea, Seoul St Marys Hosp, Div Hematol, Seoul 137701, South Korea
   Seoul Natl Univ, Coll Med, Seoul, South Korea
   Shanghai Med Univ 2, Ruijin Hosp, Shanghai, Peoples R China
   Hannover Med Sch, D-30623 Hannover, Germany
   Taipei City Hosp, Taipei, Taiwan
   Novartis Pharmaceut, E Hanover, NJ USA
   Novartis Pharma AG, Basel, Switzerland
   UCL, London, England

---

Applying the convention described for the citing publications, we assign the publication to "South Korea" which shows the highest frequency (2 out of 8) among country addresses. When the prevailing country is Italy, we can reach a higher level of detail, since we are able to define the prevailing LAU (municipality) among all the Italian addresses. To exemplify, consider the publication with DOI 10.1021/acs.inorgchem.8b02267. In the list of the 11 addresses indicated in the box below, the most frequent country is Italy. In turn, among the Italian addresses, the most frequent city is Catania, so this citing publication is located in the territory of the LAU of Catania.

---

[5] This convention has some obvious limits: a citing publication could be attributed to a given country when in fact the authors from that country did not reach a "majority" within the byline; the full counting of each of the authors' addresses distorts the result in the presence of authors with multiple affiliations; finally, the corresponding author ends up having twice as much weight as the others, for the simple fact that their affiliation appears twice in the address list. In order to evaluate the effect of such limits, we extracted a random sample of 1,000 cited publications from the dataset and, for each citing record of such publications (17,216 in all), we downloaded the author-affiliation field by means of the "Advanced Search" interface in the online WoS portal. The application of both conventions to such set of citing publications, reveals that in 96.8% of cases the "made in" country remains the same.

> Hop Prive Jacques Cartier, Inst Cardiovasc Paris, Gen Sante, Dept Cardiol, Massy, France
> CHU Cavale Blanche, Dept Cardiol, Brest, France
> Columbia Univ, Med Ctr, Dept Cardiol, New York, NY USA
> New York Presbyterian Hosp, New York, NY USA
> Univ British Columbia, Dept Cardiol, Vancouver, BC V5Z 1M9, Canada
> Univ Laval, Quebec Heart & Lung Inst, Dept Cardiol, Quebec City, PQ, Canada
> Univ Catania, Ferrarotto Hosp, Dept Cardiol, Catania, Italy
> ETNA Fdn, Catania, Italy
> Univ Turin, Div Cardiol, Citta Salute & Sci, Turin, Italy
> Imperial Coll Healthcare NHS Trust, Div Cardiol, London, England
> Univ Birmingham, Queen Elizabeth Hosp, Birmingham B15 2TH, W Midlands, England

The analysis of knowledge flows will therefore be carried out at two distinct levels:
- the international one, where the citing publications will be attributed to one and only one country on the basis of the prevalent NUTS0 code;
- the national one, in which the citing publications assigned to "Italy" are attributed to one and only one LAU (municipality) of the Italian territory, always on the basis of the prevalence criterion.

We then measure the "distances" of the citation flows, along the geodetic line[6] that joins the prevalent Italian LAU[7] of production of the aforementioned publication with:
- the capital of the citing country, for international analysis
- the citing Italian LAU, for national analysis

Overall, there are 255,399 publications in the 2010-2012 I-NCR, of which 184,177 had received at least one citation by the close of 2017. 161,680 were assigned to an Italian LAU, and had received 3,002,835 total citations from 1,800,037 unique citing publications. In turn, these citing publications were:
- from a foreign country, in 82% of cases;
- from an Italian LAU in the remaining 18% of cases.

Overall, in the dataset there are:
- 639 different Italian LAUs with a cited publication;
- 774 different Italian LAUs with a citing publication;
- 199 different countries with a citing publication.

## 4. Analysis

In this section we present the knowledge flows generated by the Italian scientific production, classified by territory, then the flows at national and international level. The impact of geographic distance on knowledge flows is studied by means of a linear regression model, controlling for other variables such as the stock of publications produced in the period under observation, in each cited and each citing territory, as a proxy for the "scientific potential" of the territories considered.

---

[6] In the literature, this method of measuring geographic distance has been adopted in Maurseth and Verspagen, 2002; Broekel and Mueller, 2018; Ahlgren, Persson and Tijssen, 2013; Jiang, Zhu, Yang, Xu, and Jun, 2018. Some scholars have instead adopted the travel time between two points (Crescenzi, Nathan, & Rodríguez-Pose, 2016; Ponds, Van Oort & Frenken, 2007).

[7] The remaining publications had more than one prevalent municipality, and have been assigned to none.

By way of example, Table 1 shows the data relating to six cited publications attributed to the LAU (municipality) of Candiolo, and to the relevant citing publications. Note that the citations from Italian publications are numerically of an order of magnitude lower than those abroad and that the same occurs for the average distances.

*Table 1: Citational flows originating from six publications attributed to the municipality of Candiolo*

|  | International analysis | | National analysis | |
|---|---|---|---|---|
| WoS_ID | Citations received | Avg_distance (km) | Citations received | Avg_distance (km) |
| 000274892500028 | 346 | 4352 | 60 | 362 |
| 000275752500008 | 102 | 5930 | 1 | 241 |
| 000276410700005 | 8 | 3407 | 3 | 210 |
| 000278246000055 | 175 | 5207 | 9 | 278 |
| 000280492100008 | 64 | 4903 | 11 | 28 |
| 000280921000051 | 34 | 4171 | 6 | 25 |

*International analysis: the citing publications are associated with only one country (if Italian then only one LAU municipality)*
*National analysis: the publications are associated with only one LAU municipality*

Table 2 presents the data aggregated by LAU2 for the cases of those municipalities with more than 2000 cited publications, grouped by macro territory (north, center, southern Italy). The values of the average distances are a little more than 4000 km at international level (with the sole exceptions of Catania and Palermo, both situated at the southern extremity of Italian national territory) and 100-200 at the national level. In the first case the average dimension is effectively transcontinental (European), in the second the distance is similar to the "diameter" of an Italian province or between two contiguous regions.

National citations are on average less than 20% of the total, with a maximum for Catania (28%) and a minimum for Verona (13%). An interesting analysis concerns the percentage of publications mentioned only by the territory to which they belong (last column of Table 2). Looking at the average values for macro-areas, it is the South that prevails in this regard (range 5%-7%), while the North and Center are similarly aligned with values never higher than 5%. The municipality with the highest percentage of exclusively local spillover publications is Catania (7.2%) while the city with the lowest percentage is Trieste (2.3%; a city in the north-eastern corner of Italy, near international border).

*Table 2: Knowledge flows from LAU municipalities with more than 2000 cited publications*

| | | International analysis | | | National analysis | | |
|---|---|---|---|---|---|---|---|
| Area | Municipality | Publications cited | Citations received * | Average distance (km) | Citations received ** | Average distance (km) | Publications cited only from the municipality |
| | Milan | 19944 | 402074 | 4295 | 66939 | 145 | 3.8% |
| | Turin | 8236 | 155319 | 4323 | 27559 | 169 | 4.9% |
| North | Bologna | 8259 | 146893 | 4167 | 27110 | 121 | 3.8% |
| | Padua | 7672 | 145349 | 4359 | 25227 | 121 | 4.3% |
| | Genoa | 4553 | 76637 | 4139 | 14462 | 140 | 4.4% |

|      |              | Publications cited | Citations received * | Average distance (km) | Citations received ** | Average distance (km) | Publications cited only from the municipality |
|------|--------------|--------------------|----------------------|-----------------------|-----------------------|-----------------------|-----------------------------------------------|
|      |              |                    | International analysis |                     | National analysis     |                       |                                               |
| Area | Municipality |                    |                      |                       |                       |                       |                                               |
|      | Trieste      | 3862               | 75220                | 4337                  | 11013                 | 178                   | 2.3%                                          |
|      | Pavia        | 3111               | 58110                | 4214                  | 10119                 | 145                   | 3.2%                                          |
|      | Trento       | 2233               | 33703                | 4351                  | 5817                  | 123                   | 4.6%                                          |
|      | Parma        | 2338               | 36933                | 4086                  | 7282                  | 129                   | 3.5%                                          |
|      | Verona       | 2070               | 42924                | 4671                  | 5564                  | 157                   | 2.7%                                          |
|        | Rome       | 23770              | 422529               | 4258                  | 76153                 | 131                   | 4.3%                                          |
|        | Pisa       | 6544               | 114872               | 4334                  | 20960                 | 118                   | 4.2%                                          |
| Center | Florence   | 5585               | 103684               | 4067                  | 18938                 | 120                   | 3.5%                                          |
|        | Perugia    | 2334               | 42178                | 4001                  | 9263                  | 115                   | 5.0%                                          |
|        | Siena      | 2240               | 35683                | 4231                  | 6622                  | 126                   | 2.7%                                          |
|       | Naples      | 7879               | 127961               | 4076                  | 29990                 | 136                   | 5.0%                                          |
|       | Bari        | 3687               | 60423                | 4199                  | 12596                 | 217                   | 5.6%                                          |
| South | Catania     | 3321               | 46020                | 3696                  | 12886                 | 211                   | 7.2%                                          |
|       | Palermo     | 3116               | 44663                | 3934                  | 10776                 | 197                   | 6.2%                                          |
|       | Messina     | 2029               | 25678                | 3777                  | 7036                  | 191                   | 7.0%                                          |

*\* Including those from Italian citing publications*
*\*\* From Italian citing publications only*

As mentioned above, in order to quantify the influence of the "geographic distance" factor on the citation flows observable through our dataset, we use an econometric model, in particular a gravity model, specified in this manner in the case of national analysis:

$$C_{ij} = k \cdot \frac{M_i^\alpha M_j^\beta}{d_{ij}^\gamma}$$

[1]

with:
$C_{ij}$ = number of citations to publications from the cited LAU (municipality) $i$ by the publications of the citing LAU $j$
$k$ = constant
$M_i$ = number of publications produced in total by cited LAU $i$ in the 2010-2012 period
$M_j$ = number of publications produced in total by citing LAU $j$ in the 2010-2017 period
$d_{ij}$ = geodetic distance between cited LAU $i$ and citing LAU $j$

For the international analysis, the following apply:
$C_{ij}$ indicates the number of citations to publications from the cited LAU $i$ by the publications of citing country $j$
$M_j$ is the prevalent country $j$
$d_{ij}$ is the distance between cited LAU $i$ and the capital of the citing country $j$

Applying a logarithmic transformation to all variables of equation [1], we obtain:

$$\ln(C_{ij}) = \ln(k) + \alpha\ln(M_i) + \beta\ln(M_j) - \gamma\ln(d_{ij}) + \varepsilon \qquad [2]$$

The coefficients of a log-log model represent the elasticity of the Y dependent variable with respect to the X independent variable. For example, for the distance variable ($d_{ij}$) an elasticity of one ($\gamma = 1$) indicates that a 1% increase in the distance is associated with a 1% decrease in citations exchanged, on average.

Table 3 presents the results of the ordinary least squares (OLS) regression applied to the national analysis.

*Table 3: Results of the OLS regression for the national analysis*

| Variable | Coeff. | | Robust Std Err. |
|---|---|---|---|
| $M_i$ | 0.437 | *** | 0.006 |
| $M_j$ | 0.437 | *** | 0.006 |
| $d_{ij}$ | 0.474 | *** | 0.011 |
| k | -1.773 | *** | 0.073 |
| $R^2$ | 0.549 | | |
| Obs | 10786 | | |

*Significance level: *** 0.01, ** 0.05, * 0.1.*

From Table 3 we can observe that the coefficients of the three independent variables are very similar in absolute value. The two coefficients of masses are significant, positive and equal to 0.437, which seems to be a logical outcome. Instead, we see slightly greater elasticity for the distance variable, at 0.474; therefore, for the same masses, a 1% increase in the distance between territories is associated with a 0.474% decrease in citations exchanged, on average between territories. This indicates an obvious effect of geographic proximity on knowledge flows. At the aggregate level, the value of the coefficient γ suggests that distance still matters in science, even in the case of the analysis being limited to knowledge flows between subjects within a single national contest - a finding in line with previous literature.

Table 4 presents the results from the international analysis. Given the geographic distances involved, we decided to further subdivide this analysis by distinguishing between continental flows (i.e. referred to citations from European countries) and intercontinental flows (i.e. referred to citations from non-European countries).

*Table 4: Results of the OLS regression for the international analysis*

| | Europe dataset | | | Extra-Europe dataset | | |
|---|---|---|---|---|---|---|
| Variable | Coeff. | | Robust Std Err. | Coeff. | | Robust Std Err. |
| $M_i$ | 0.762 | *** | 0.005 | 0.701 | *** | 0.007 |
| $M_j$ | 0.820 | *** | 0.008 | 0.781 | *** | 0.009 |
| $d_{ij}$ | 0.423 | *** | 0.018 | -0.051 | *** | 0.018 |
| Const | -8.185 | *** | 0.168 | -11.166 | *** | 0.185 |
| $R^2$ | 0.758 | | | 0.726 | | |
| Obs | 7895 | | | 8163 | | |

*Significance level: *** 0.01, ** 0.05, * 0.1. Robust Standard errors in brackets*
****; **; * are statistically significant at 1%; 5%; 10% levels*

Comparing these new results with those of Table 3, it can be observed that, moving from the national to the international level, the coefficients of the regressors $M_i$ and $M_j$ (although always quite similar) increase in absolute value. On the other hand, with regard to the distance variable, in the analysis of flows limited to Europe, the coefficient γ is slightly lower (0.423) than that recorded in the national analysis (0.474). On the other hand, observing the citation flows from extra-Europe countries, the coefficient assumes a negative value close to zero (-0.051), indicating that on an intercontinental scale the geographic distance becomes a negligible factor in understanding the dynamics of knowledge flows. Looking at the list of the countries citing Italian publications, in the top of the rank we find USA followed by China, Japan, Canada, Australia, India, South Korea, Brazil, Taiwan and Iran, i.e countries at science frontier and very big in terms of size of their science system. So, it is somewhat expected that in the hierarchy of these countries, the geographical distance to Italy has no influence, also given the presence of oceans in between in most cases.

On the basis of these results, we conducted an ad hoc examination of the distance regressor for the analysis of national flows, as seen below. Similar to what Belenzon and Schankerman (2013) did in their collaborations study, to analyze nonlinear effects of distance we use a set of four dummy variables for intervals of distance. As indicated in Table 5, the reference category is 0-50 km, which might be interpreted as a "metropolitan effect".

*Table 5: Codification of dummy variables for the analysis of national flows*

| Dummy code | Distance range (km) | Dummies vector |
|---|---|---|
|  | 0-50 | 0;0;0 |
| a | 50-400 | 1;0;0 |
| b | 400-800 | 0;1;0 |
| c | 800-1200 | 0;0;1 |

The results of the OLS regression reported in Table 6 show that, with respect to the metropolitan area, for all the bands considered, the coefficients are statistically significant and increasing, indicating a progressive decrease in knowledge flows that is all the more marked the further away the territories are.

*Table 6: Comparison of the results of the OLS regression for the national dataset, with two different specifications of distance*

| Variable | Coeff. | | Robust Std Err. |
|---|---|---|---|
| $M_i$ | 0.419 | *** | 0.006 |
| $M_j$ | 0.422 | *** | 0.006 |
| dummy_a | 1.679 | *** | 0.063 |
| dummy_b | 1.836 | *** | 0.064 |
| dummy_c | 1.944 | *** | 0.070 |
| Const | -2.540 | *** | 0.080 |
| $R^2$ | 0.506 | | |
| Obs | 10786 | | |

## 5. Conclusions

This paper analyses the influence of geographic distance on the knowledge flows between producers (of cited articles) and users (by means of the citing articles). It differs from all previous ones on the theme for the overall sectoral breadth and much greater size of the field of observation. The pairs of localities connected by a citation bond were considered as the units of analysis.

The use of a gravitational model estimated by OLS, controlling for the mass $M_i$ and $M_j$ of the territories citing and cited, shows a decrease in domestic knowledge flows between territories as their distance increases. This confirms the presence of an effect of geographic proximity, serving as the basis of any localization phenomena, in which direct personal interaction plays an important role in knowledge flows. A further analysis, which distinguished the continental flows (referred to citations from European countries) from the intercontinental flows (referred to citations from non-European countries), shows that the weight of the geographic factor on the knowledge flow dynamics decreases progressively, ultimately disappearing with the non-European countries.

However, there are two circumstances that we expect would attenuate the importance of geographic proximity: (i) the nature of knowledge conveyed by scientific publications is "encoded", which means that it should be less sensitive and influenced by geographic proximity than "tacit" knowledge; (ii) the widespread and pervasive use of IT, in which encoded is the most easily transmitted knowledge, could facilitate the almost instantaneous dissemination of information.

The fact that the geographic factor is still present and statistically significant in the analysis carried out may be attributable to a number of determinants such as, for example: i) the specification of the model adopted and the variables considered (although those identified are the most plausible from a bibliometric point of view); ii) a significant incidence of self-citations; iii) the presence of a series of latent variables not explained, classifiable as "social factors" (links between mentors and students, belonging to the same scientific school in a field; an asymmetry of the citation process in favor of papers published in prestigious journals or prestigious scientists, the presence of a "country" effect especially with regard to publications cited at the international level).

These are all aspects related to the nature of the citation process, in which the citations not only reflect the attribution of scientific credit but also include other external components of a "social" nature, and this could then generate a bias in favor of the geographic factor, especially at the domestic level.

The generalization of data to other national contexts must be applied with due caution, for several reasons, first of all the geographic conformation of Italy and the context relative to some of the most scientifically active countries.

The authors propose to deepen the topic in a future work, analyzing and comparing knowledge diffusion across scientific fields. In particular, it will be important to observe if and how the effect of geographic distance varies between fields, due to the different correspondence of sectoral research to local needs and/or the different share of tacit component in the information to be transmitted: the less it can be encoded, the more important should be the personal relationship underlying the exchange, which would make knowledge flow more sensitive to the geographic factor.

A further study that the authors are about to complete concerns the effect of time on the factor of geographic proximity, i.e. whether over the years after production of the knowledge, geographic proximity still remains a determining factor for its dissemination. In that work, we also analyse the bias caused by self-citations when measuring the effect of geographic proximity in knowledge flows. When self-citations are controlled for, first results show that the importance of geographical proximity decreases, especially when domestic flows are considered. Furthermore, the share of self-citations (and their bias) is significantly decreasing when citations are traced for longer periods. These results are aligned with those of Wuestman, Hoekman, and Frenken (2019) in the life sciences and medicine.